\documentclass[aps,showpacs,pra,preprint]{revtex4}
\usepackage{mathrsfs} 
\usepackage{amssymb}

\usepackage{graphicx}
\usepackage{dcolumn}
\usepackage{bm}

\begin{document}

\title{Entanglement dynamics of photon pairs emitted from quantum dot}
\author{Yang Zou, Ming Gong, Chuan-Feng Li$\footnote{
email: cfli@ustc.edu.cn}$, Geng Chen, Jian-Shun Tang and Guang-Can Guo} \affiliation{Key Laboratory of Quantum Information, University of Science and Technology of
China, CAS, Hefei, 230026, People's Republic of China}
\date{\today }


\begin{abstract}
We present a model that describes states of photon pairs, which have been generated by biexciton cascade decays of self-assembled quantum dots, use of which yields
finding that agrees well with experimental result. Furthermore, we calculate the concurrence and determine the temperature behavior associated with so-called
entanglement sudden death, that prevents quantum dots emitting entangled photon pairs at raised temperatures. The relationship between the fine structure splitting
and the sudden death temperature is also provided.
\end{abstract}

\maketitle

Entanglement, the intriguing correlations of quantum systems, has found extensive application in the areas of quantum computation \cite{Knill2001nature} and quantum
communication \cite{Gisin2002rmp}. Although this fundamental feature has attracted a lot of attention since the early years of quantum mechanics, it is still far
from being completely understood. In recent times, experimentation has found that entanglement within a two-qubit system can disappear abruptly within a finite time
during the decoherence process \cite{Yu2004prl,Yu2006oc,Eberly2007science}. This phenomenon, termed ``entanglement sudden death" (ESD), has been demonstrated in
optical systems under rather special artificial environments \cite{Almeida2007science,Xu2009prl}.

The generation of entangled photon pairs by using a biexciton cascade process in a single quantum dot (QD) was first proposed by Benson {\it et al.}
\cite{Benson2000prl}, and has been recently realized by several groups \cite{Akopian2006prl,Srevenson2006nature}. In this process, two electrons and two holes are
initially generated in the QD (biexciton XX), and then a biexciton photon $H_{XX}$ or $V_{XX}$ is emitted as the dot decays to an exciton ($X$) state by recombining
one electron and one hole. The polarization of the biexciton photon is either horizontal ($H$) or vertical ($V$), in accord with the decay into the exciton state
$X_H$ or $X_V$, respectively. After a time delay $\tau$, the other electron and hole recombine to emit an exciton photon $H_X$ or $V_X$ with the same polarization
as that of the earlier biexciton photon. When the two recombination paths ($H$ or $V$)are indistinguishable in frequency, the output is the maximum entangled Bell
state $|\Phi\rangle = (|HH\rangle + |VV\rangle)/\sqrt{2}$.

Unfortunately, in QDs, due to the underlying atomic symmetry being $C_{2v}$ rather than $C_{4v}$ \cite{Bayer2002prb,bester2005prb}, the intermediate two exciton
state has a small splitting of about several tens of $\mu$eV, much larger than the intrinsic width of the emission line ($\sim$ $\mu$eV). This splitting, labeled as
$S$ in Figure\,1, has conventionally been called the ``fine structure splitting" (FSS). Due to this FSS, the actual output is not a pure polarization entangled
state, but a state with both polarization and frequency entangled, along with a phase difference between $H$ and $V$ \cite{Stevenson2008prl},
\begin{equation}
|\Psi\rangle = \frac{1}{\sqrt{2}}(|H_{XX}H_X\rangle + e^{iS\tau}|V_{XX}V_X\rangle).\label{idealstate}
\end{equation}
If the frequency is traced over, this state becomes just a mixed state with classical correlations; only when $S \sim 0$ can we achieve an entangled state.

In experiments, there are many methods to tune the FSS, such as by annealing \cite{Young2005prb}, exerting uniaxial stress \cite{Seidl2006apl}, or applying a
magnetic field \cite{Stevenson2006prb} or an electric field \cite{Marcet2010apl}. Meanwhile, various other experiments are designed to erase the frequency
information by filtering the photons spectrally \cite{Akopian2006prl} or selecting the exciton ($X$) photon within only a small emission delay relative to the
biexciton ($XX$) photon \cite{Stevenson2008prl}, by which entanglement is greatly enhanced.

In this work, taking the acoustic phonon-assisted transition process into account, we investigated entanglement for photon pairs generated from a QD. A similar ESD,
which depends on temperature and FSS, has been found in this process.
\begin{figure}[b]
\centering
\includegraphics[width= 2.7in]{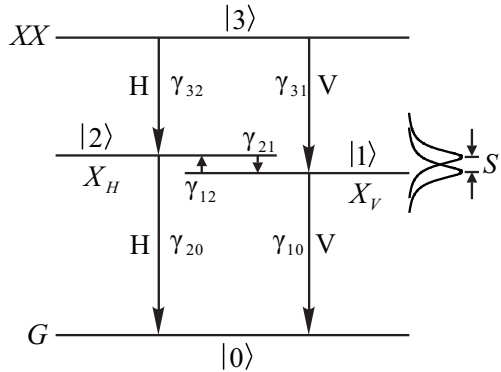}
\caption{Energy level schematic of the biexciton cascade process, the ground state $G$ ($|0\rangle$), the two linear polarized exciton state $X_{H}$ ($|2\rangle$)
and $X_{V}$ ($|1\rangle$) and the biexciton state $XX$ ($|3\rangle$). The spontaneous emission process is denoted by $\gamma_{ij}$ ($\gamma_{12}$ and $\gamma_{21}$
are phonon assisted transition rates).}\label{fourlevel}
\end{figure}

A schematic of the biexciton cascade process in a QD is presented in Fig.\,\ref{fourlevel}, where the vacuum state $(G)$, the two intermediate exciton states
($X_{V}$ and $X_{H}$), and the biexciton state $(XX)$ are labeled respectively as $|0\rangle \sim |3\rangle$ for convenience.

The QD is initially excited to the biexciton state by a short-pulsed laser, and subsequently evolves freely. Considering the phonon assisted process, the density
matrix $(\hat{\rho})$ of this four-level system can be derived using the master equation ($\hbar = 1$) \cite{Wang2005prl},
\begin{eqnarray}
\dot{\hat{\rho}}=-i[\hat{H_0},\hat{\rho}]+L(\hat{\rho}),\label{mastereq}
\end{eqnarray}
where
\begin{eqnarray}
\hat{H_{0}}=\sum_{i=0}^3 \omega_{i}|i\rangle\langle i|.
\end{eqnarray}
The dissipation term in the Lindblad form is described by
\begin{eqnarray}
L(\hat{\rho}) & = & \frac{1}{2}[ \gamma_{32}(|2\rangle\langle 3|\hat{\rho}|3\rangle\langle 2|-|3\rangle\langle 3|\hat{\rho}-\hat{\rho}|3\rangle\langle
3|)\nonumber\\
& & +\gamma_{31}(|1\rangle\langle 3|\hat{\rho}|3\rangle\langle 1|-|3\rangle\langle 3|\hat{\rho}-\hat{\rho}|3\rangle\langle 3|)\nonumber\\
& & +\gamma_{20}(|0\rangle\langle 2|\hat{\rho}|2\rangle\langle 0|-|2\rangle\langle 2|\hat{\rho}-\hat{\rho}|2\rangle\langle 2|)\nonumber\\
& & +\gamma_{10}(|0\rangle\langle 1|\hat{\rho}|1\rangle\langle 0|-|1\rangle\langle 1|\hat{\rho}-\hat{\rho}|1\rangle\langle 1|)\nonumber\\
& & +\gamma_{21}(|1\rangle\langle 2|\hat{\rho}|2\rangle\langle 1|-|2\rangle\langle 2|\hat{\rho}-\hat{\rho}|2\rangle\langle 2|)\nonumber\\
& & +\gamma_{12}(|2\rangle\langle 1|\hat{\rho}|1\rangle\langle 2|-|1\rangle\langle 1|\hat{\rho}-\hat{\rho}|1\rangle\langle 1|)],
\end{eqnarray}
where $\gamma_{32}$, $\gamma_{31}$, $\gamma_{20}$, $\gamma_{10}$ are the spontaneous emission rates, while $\gamma_{21}$ and $\gamma_{12}$ denote the phonon
assisted transition rates between $|1\rangle$ and $|2\rangle$. The phonon absorption rate of the state $|1\rangle$ is $\gamma_{12}=\kappa N_{B}$ and the emission
rate of $|2\rangle$ is $\gamma_{21}=\kappa(N_{B}+1)$, where $\kappa$ is the phonon-QD interaction rate, which is approximately proportional to the cube of the
energy splitting $S$ \cite{Shen2007prb}, and $N_{B}$ represents the Bose distribution function of a phonon with energy $S$, $N_{B}=[\exp(S/k_{B}T)-1]^{-1}$.

This emission process produces radiation in a mixed state, from which the two-photon coincidence measurements single out the polarization density matrix of the
photon pair $\hat{\rho}_{\text{pol}}$, which is projected onto the subspace spanned by the four basis $\{|H_{1}H_{2}\rangle$, $|H_{1}V_{2}\rangle$,
$|V_{1}H_{2}\rangle$, $|V_{1}V_{2}\rangle\}$. Its matrix elements, experimentally reconstructed by means of quantum tomography, theoretically read as
\cite{Troiani2006prb}
\begin{eqnarray}
\left\langle\mu_{1}\nu_{2}\right|\hat{\rho}_{\text{pol}}\left|\xi_{1}\zeta_{2}\right\rangle & = & \mathcal
{A}\int\limits_{t_{m}}^{t_{M}}dt\int\limits_{t^\prime_{m}}^{t^\prime_{M}}dt^\prime\nonumber\\
&\times& \left\langle\sigma^\dag_{\mu_{1}}(t)\sigma^\dag_{\nu_{2}}(t^\prime)\sigma_{\zeta_{2}}(t^\prime)\sigma_{\xi_{1}}(t)\right\rangle,\label{correlation}
\end{eqnarray}
where $(\mu,\nu,\xi,\zeta) \in \{H,V\} $, and the $\sigma$s are the dipole transition operators, $\sigma_{_{H_1}}\equiv|3\rangle\langle2|$,
$\sigma_{_{V_1}}\equiv|3\rangle\langle1|$, $\sigma_{_{H_1}}\equiv|2\rangle\langle0|$, $\sigma_{_{V_2}}\equiv|1\rangle\langle0|$. $\mathcal {A}$ is a normalization
factor to ensure that the trace of $\hat\rho_{\text{pol}}$ is unitary. The limits $t_{m}$ and $t_{M}$ ($t^\prime_{m}$ and $t^\prime_{M}$) defined the temporal
window related to the detection of the photons from the biexciton cascade process. The emission delay of the exciton photon relative to the biexciton photon, see in
Eq.\,(\ref{idealstate}), is $\tau\equiv t^\prime-t$. All these correlation functions are computed by applying the quantum regression theorem \cite{Lax1968pr}.

Ideally, the polarization density matrix calculated from Eqs.\,(\ref{mastereq}) - (\ref{correlation}) has the form
\begin{eqnarray}
\hat{\rho}_{\text{pol}}=\left(\begin{array}{cccc}
  \rho_{11} & 0 & 0 & \rho_{14} \\
  0 & \rho_{22} & 0 & 0 \\
  0 & 0 & \rho_{33} & 0 \\
  \rho_{41} & 0 & 0 & \rho_{44}
  \end{array}\right).
\end{eqnarray}
All non-diagonal terms except $\rho_{14}$ and $\rho_{41}$ are zero, and the physical reason for this is that the Hamiltonian in Eq.(\ref{mastereq}) only couples
states that share the same excitation labels $H$ and $V$.

In realistic situations, aside from phonon-induced spin scattering, the system may be affected by some other decoherence processes, such as decay-path
distinguishability, background noise, cross dephasing and pure dephasing. However, cross-dephased light has been shown to be weak \cite{Hudson2007prl}, and photon
pair emission in a QD is robust against single photon decoherence being typically limited by pure dephasing \cite{Hudson2007prl,Stevenson2008prl}. Thus we can
ignore cross and pure dephasings in the calculation. Taking the other factors into account, the total density matrix ($\hat{\rho}_{\text{tot}}$) of the photon pair
is divided into three parts,
\begin{eqnarray}
\hat{\rho}_{\text{tot}} = \frac{1}{1+g}(\eta\hat{\rho}_{\text{pol}} + (1-\eta)\hat{\rho}_{\text{noc}} + g\hat{\rho}_{\text{noise}} \label{rhot}),
\end{eqnarray}
where $\hat\rho_{\text{noc}}$ arises from the FSS-induced distinguishability between the two decay paths, representing the non-overlapping part of the emission
lines of excitons $X_H$ and $X_V$, as shown in Fig.\,\ref{fourlevel}. It describes those photon pairs which can be distinguished from the spectra domain, and have
only a classical correlation without a phase relationship; hence, the non-diagonal elements of $\hat\rho_{\text{noc}}$ are all zero, while the diagonal elements
evolve in the same manner as $\hat{\rho}_{\text{pol}}$. The third term $\hat\rho_{\text{noise}}$, which describes the background noise, is set as an identity
matrix.

\begin{figure}[tbph]
\centering
\includegraphics[width= 3in]{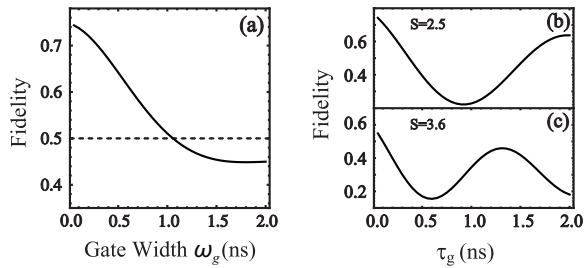}
\caption{(a) Fidelity as a function of gate width $w_g$ ($\tau_{g}=0$) at $S = 2.5$ $\mu$eV. Dashed horizontal line represents the limit for classical behavior. (b)
and (c) present the fidelity as a function of  $\tau_{g}$ with gate width $w_g = 0.5$ ns for $S=2.5$ and $S=3.6$  $\mu$eV, respectively. }\label{fitresult}
\end{figure}
In Ref.\,[\onlinecite{Stevenson2008prl}], Stevenson {\it et al.} detected the photon pairs with delays $\tau$ in the range $\tau_{g}\leq\tau\leq(\tau_{g}+w_{g})$ by
applying a single timing gate. In this case, the non-diagonal term $\rho_{14}$ ($\rho_{41} = \rho_{14}^\ast$) has the form of an exponential decay
\begin{eqnarray}
\rho_{14} && \propto \int\limits_{\tau_{g}}^{\tau_{g}+w_g}  d\tau
\exp{[-\Gamma\tau/2+iS\tau]}  \nonumber \\
&&  \propto {e^{(iS-\Gamma/2)\tau_{g}} \over iS-\Gamma/2} (e^{(iS-\Gamma/2)w_g} - 1), \label{eqrho14}
\end{eqnarray}
where $\Gamma = \gamma_{20}+\gamma_{10}+\gamma_{12}+\gamma_{21}$. Our results shown in Fig.\,\ref{fitresult} can explain their experimental behaviors as being a
consequence of a maximum fidelity 0.73 of Bell state $|\Phi\rangle$ obtained when gate width was $49$\,ps. Also, a marked oscillation in fidelity as a function of
$\tau_{g}$ was observed, when the same parameters as that in Ref.\cite{Stevenson2008prl} are chosen: $\gamma_{20}=\gamma_{10}=1.3$ ns$^{-1}$,
$\gamma_{32}=\gamma_{31}=1.8$ ns$^{-1}$ (that corresponds to the inverses of the exciton and biexciton lifetimes within a QD), and temperature $T = 10$\,K, a value
for $\eta$, the overlap part of the two exciton spectra, of $0.91$ when $S = 2.5$\,$\mu$eV. To fit the experimental results, we chose $g=0.45$. When we fix the gate
width in Eq.\,(\ref{eqrho14}), the fidelity as a function of $\tau_{g}$ has a sinusoidal envelop, as evident in Fig.\,2\,(b) for $S = 2.5$\,$\mu$eV and (c) for $S =
3.6$\,$\mu$eV. The frequency of the oscillation increases as the FSS widens. Furthermore, we fix $\tau_{g}= 0$, and change the gate width $w_g$, in the long time
scale, the integration of Eq.\,(\ref{eqrho14}) approaches zero.
\begin{figure}[b] \centering
\includegraphics[width=2.5in]{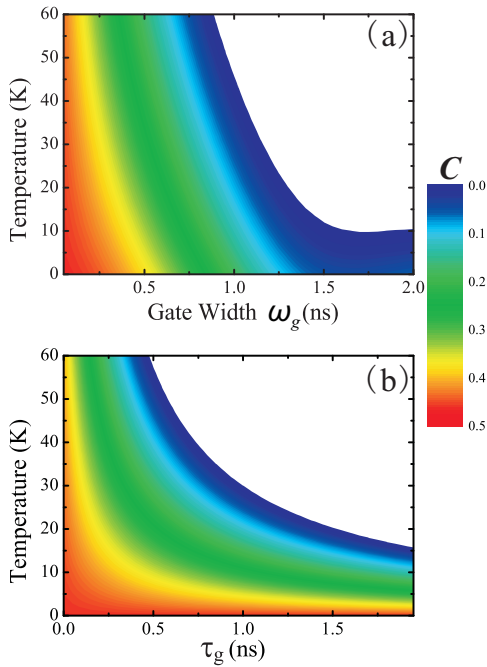}
\caption{(Color online). Concurrence as a function of temperature and (a) gate width $w_g$ ($\tau_g=0$), (b) delay time $\tau_g$ ($w_g=0.1$). The FSS is
$2.5$\,$\mu$eV. }\label{wtg}
\end{figure}
Consequently, the fidelity is below 0.5 (the maximum achievable fidelity for an unpolarized classical state). In contrast, when the gate width is shorter than
$1$\,{ns}, the fidelity can increase above the classical limit, as shown in Fig.\,\ref{fitresult}\,(a).

However, fidelity is insufficient in describing the entanglement properties of the photon pair. To quantify this, a widely used form of entanglement is Wootters'
concurrence \cite{Wootters1998prl}, which is well defined. The concurrence $C$ of a two-particle system is defined as
\begin{equation}
C(\hat\rho_{\text{tot}})= \text{max}\{0,\sqrt{\mathstrut \lambda_1}-\sqrt{\mathstrut \lambda_2}-\sqrt{\mathstrut \lambda_3}-\sqrt{\mathstrut \lambda_4}\},
\end{equation}
where $\lambda_i$ are the eigenvalues of the matrix $\hat\rho_{\text{tot}}(\sigma_2\otimes\sigma_2) \hat\rho_{\text{tot}}^\ast(\sigma_2\otimes\sigma_2)$ arranged in
decreasing order, and $\hat\rho_{\text{tot}}$ is defined in Eq.\,(\ref{rhot}).

The result is presented in Fig.\,\ref{wtg} where the FSS energy is $2.5$\,$\mu$eV. As temperature rises, concurrence decreases due to strong phonon induced
scattering processes. In Fig.\,\ref{wtg}\,(a), entanglement declines rapidly for large gate widths, that can be ascribed to obtaining more which-path information
from the FSS in the energy domain. This is because the Fourier transform of a truncated exponential decay results in a broad natural linewidth of the post-selected
photons. Long time transition processes between two exciton states also reduce entanglement, as shown in Fig.\,\ref{wtg}\,(b) where $w_g$ is fixed at $0.1$\,ns.

Under both situations, ESD is observed at high temperatures. It is worth noticing that this ESD is quite different from those in quantum optics experiments
\cite{Almeida2007science,Xu2009prl}, where one can alter the initial state to prevent the system from undergoing ESD. In a QD system, the ESD is mainly due to the
intrinsic phonon assisted transition process between $|1\rangle$ and $|2\rangle$. Thus, we cannot change the initial state, but FSS can be tuned in experimnet, and
we can show that the ESD phenomenon will disappear when the FSS is small and the temperature is low.

\begin{figure}[top]
\centering
\includegraphics[width=3.3in]{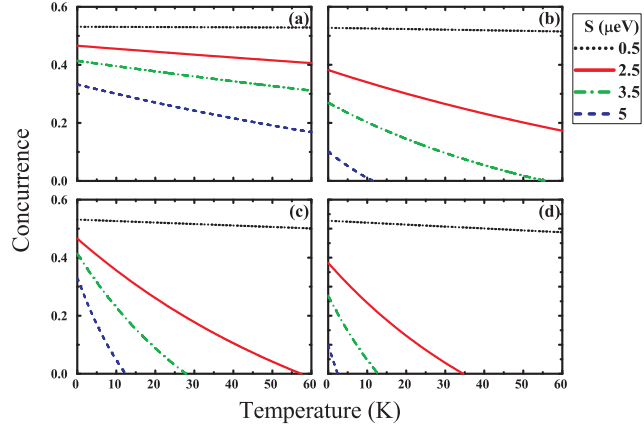}
\caption{(Color online). Concurrence as a function of temperature for different FSS energies. The four kinds of line shapes represent that S is $0.5$, $2.5$, $3.5$,
and $5$\,$\mu$eV respectively. (a) $w_g = 0.1$\,ns and $\tau_g = 0$\,ns. (b) $w_g = 0.5$\,ns and $\tau_g = 0$\,ns. (c) $w_g = 0.1$\,ns and $\tau_g = 0.5$\,ns. (d)
$w_g = 0.5$\,ns and $\tau_g = 0.5$\,ns.}\label{CT}
\end{figure}

Further, the influence of temperature and FSS are investigated, as shown in Fig.\,\ref{CT}. The photon pair generated by a QD with large FSS has lower degree of
entanglement, that also falls off rapidly. However, for a QD with small FSS, such as $S = 0.5$\,$\mu$eV, the concurrence decays slightly with temperature, because a
tiny phonon assisting emission rate results in a small transition rate in this small-FSS system. Comparing Fig.\,\ref{CT}\,(a) and (c), for which the evolution time
is the only differing parameter, entanglement at low temperature is found to be almost the same for every FSS, although the decreasing rates are apparently faster
for longer evolution times (Fig.\,\ref{CT}\,(c)) and ESD temperatures are lower for larger FSSs. In Fig.\,\ref{CT}\,(b), the degree of entanglement is already small
at low temperature due to its large gate width, and the reduction rate is faster than that in (a), a fact attributable to a system evolving within the time duration
of the gate window. In Fig.\,\ref{CT}\,(d), the influence of long evolution times and the gate widths are taken into account, so the concurrence initial values at
low temperatures are small, and reduction rates are the fastest.

\begin{figure}[top]
\centering
\includegraphics[width=2.5in]{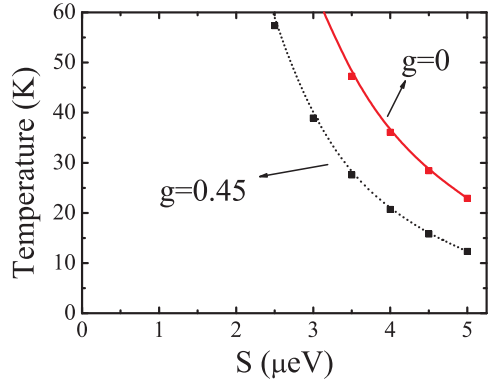}
\caption{(Color online). The relationship between the sudden death temperature and splitting S. Gate width $w_{g}=0.1$\,ns. }
\end{figure}

Finally, we show that the observed ESD is independent of background light. In Fig.\,5, we present the relationship of sudden death temperature as a function of
splitting $S$ for different $g$, for which we choose $\tau_g=0.5$\,ns and $w_g = 0.1$\,ns. Even in the ideal case, i.e. $g=0$, we still observe ESD. The ESD
temperature decreases fast with respect to FSS, owning to the fact that for large FSS, not only is the transition rate between the two exciton states fast, but also
the initial degree of entanglement is very small, due to the distinguished part $\hat{\rho}_{\text{noc}}$ being dominant in the total density matrix
$\hat{\rho}_{\text{tot}}$. This indicates that a large-FSS QD cannot generate good entangled photon pairs even at low temperatures.

In conclusion, we have calculated the concurrence of photon pairs emitted from quantum dots and considered how temperature and fine structure splitting affect
entanglement between a photon pair. An enhancement of entanglement sudden death with temperature has been found. This obviously prevents quantum dots emitting
entangled photon pairs at high temperatures. We also point out that this entanglement sudden death is a characteristic of quantum dot systems.

This work was supported by National Fundamental Research Program, and National Natural Science Foundation of China (Grant Nos. 60121503, 10874162 and 10734060).

\end{document}